\documentclass[onecolumn,aps,prd,superscriptaddress,nofootinbib,11pt]{revtex4-2}
\usepackage[utf8]{inputenc}
\usepackage[T1]{fontenc}
\usepackage[tbtags]{amsmath}
\usepackage{amssymb}
\usepackage{dsfont}
\usepackage{graphicx}
\usepackage{booktabs}
\usepackage{xcolor}
\definecolor{vfiveblue}{RGB}{0,65,170}
\newif\ifvfivehighlight
\vfivehighlightfalse
\newcommand{\vfive}[1]{\ifvfivehighlight\textcolor{vfiveblue}{#1}\else#1\fi}
\usepackage{hyperref}
\hypersetup{colorlinks=true,citecolor=black,linkcolor=black,urlcolor=black}

\newcommand{\re}{\mathrm{Re}\,}
\newcommand{\im}{\mathrm{Im}\,}
\newcommand{\sech}{\mathrm{sech}}
\newcommand{\one}{\mathds{1}}
\newcommand{\BCHSH}{\mathcal B_{\rm CHSH}}
\newcommand{\Rr}{\mathcal R}
\allowdisplaybreaks

\begin{document}

\title{Majorana bilinears, compact modular seeds and the approach to the Tsirelson bound}

\author{M.~S.~Guimaraes}
\email{msguimaraes@uerj.br}
\affiliation{UERJ -- Universidade do Estado do Rio de Janeiro, Instituto de F\'isica -- Departamento de F\'isica Te\'orica, Rua S\~ao Francisco Xavier 524, 20550-013, Maracan\~a, Rio de Janeiro, Brazil}

\author{I.~Roditi}
\email{roditi@cbpf.br}
\affiliation{CBPF -- Centro Brasileiro de Pesquisas F\'isicas, Rua Dr.~Xavier Sigaud 150, 22290-180, Rio de Janeiro, Brazil}
\affiliation{Institute for Theoretical Physics, ETH Z\"urich, 8093 Z\"urich, Switzerland}

\author{S.~P.~Sorella}
\email{silvio.sorella@fis.uerj.br}
\affiliation{UERJ -- Universidade do Estado do Rio de Janeiro, Instituto de F\'isica -- Departamento de F\'isica Te\'orica, Rua S\~ao Francisco Xavier 524, 20550-013, Maracan\~a, Rio de Janeiro, Brazil}

\begin{abstract}
In this work we study the Bell-CHSH inequality in the vacuum of the free massive Majorana field in $1+1$ dimensions, taking as the dichotomic observables of the two parties even bilinears of smeared fields localized in the right and in the left Rindler wedges. The test functions are generated from seeds in the modular-frequency variable through the Tomita-Takesaki operator of the right wedge and, on the side of the left wedge, through its adjoint multiplied by $i$, which accounts for twisted duality. For seeds supported at positive modular frequency the one-particle overlaps between the two parties do not depend on the modular weights, and we obtain the Bell parameter in closed form. When the seeds are phase rotations of a single compact bump, the bilinears generate a copy of the Pauli algebra in each wedge, and we compute exactly the correlation matrix of the vacuum restricted to the two copies. The optimal settings violate the classical bound up to spectral widths more than twice as large as those reached with planar settings. We also show that the Tsirelson bound is never attained in the wedge vacuum for any choice of dichotomic observables, and that approaching it forces the modular spectral weight of the Bell vectors to concentrate at the \vfive{spectral point $\lambda=1$} of the modular operator. This establishes in a precise way how the even sector of the free Majorana field realizes the Summers-Werner mechanism.
\end{abstract}

\maketitle

\section{Introduction}\label{sec:intro}

The inequality derived by Bell~\cite{Bell:1964kc}, in the form given to it by Clauser, Horne, Shimony and Holt~\cite{Clauser:1969ny}, bounds by $2$ the correlations that any local hidden-variable description can assign to two parties performing dichotomic measurements. Quantum theory allows values up to $2\sqrt2$, as shown by Tsirelson~\cite{Cirelson:1980ry}. In relativistic quantum field theory the question of how close the vacuum can come to this bound was settled by Summers and Werner~\cite{Summers:1985pzz,Summers:1987fn,Summers:1987squ,Summers:1987ze}. They proved that the vacuum violates the Bell-CHSH inequality maximally for observables localized in complementary wedges, for free Bose and for free Fermi fields, and that maximal violation is in fact generic under the standard assumptions of the algebraic approach. An early field-theoretic analysis of the same question is due to Landau~\cite{Landau:1987}. The argument of Summers and Werner rests on the Tomita-Takesaki theory of the wedge algebras, whose modular group was identified by Bisognano and Wichmann~\cite{Bisognano:1975ih,Bisognano:1976za} with the Lorentz boosts preserving the wedge. It also relies on the possibility of choosing one-particle vectors whose spectral weight sits near the point $\lambda=1$ of the spectrum of the modular operator. The same identification underlies the thermal character of the vacuum restricted to a wedge, the algebraic counterpart of the effect found by Unruh~\cite{Unruh:1976db}. The modular localization of Brunetti, Guido and Longo~\cite{Brunetti:2002nt}, reviewed by Guido in~\cite{Guido:2008jk}, turns it into a practical tool for constructing localized vectors without writing down test functions in spacetime.

In a series of recent works we have tried to make this mechanism explicit in concrete models. For the massive scalar field in $1+1$ dimensions, we used in~\cite{Caribe:2026mam} the Tomita-Takesaki construction of wedge-localized one-particle vectors to examine which Bell operators, Weyl operators among them, approach the Tsirelson bound. We showed in~\cite{Caribe:2026xoa} that the vertex operators of a chiral boson provide bounded dichotomic operators whose correlator can be brought arbitrarily close to it, and we discussed finite Weyl polynomials and a broader class of bounded Hermitian operators in~\cite{Caribe:2026pje,Guimaraes:2025vfu}. For the massive Majorana field we gave in~\cite{Caribe:2026tfe} a rapidity-space realization of the Summers-Werner construction, in which the vacuum correlator of normalized smeared fields reduces to a single spectral weight for the modular operator. Analytic families then approach the Tsirelson bound as that weight concentrates near zero modular frequency. Fermions bring a difficulty of their own, since smeared Majorana fields localized in complementary wedges anticommute. Bell operators built directly from them require the twisted version of Haag duality, established by Foit~\cite{Foit:1983pbi} for free Fermi fields and proved for the self-dual CAR algebra in every Fock representation by Baumg\"artel, Jurke and Lled\'o~\cite{Baumgartel:1999vc}. Even operators avoid the twist altogether, because they commute in the ordinary sense with everything localized in the complementary region.

The even sector is also the one singled out by fermionic quantum information. Ba\~nuls, Cirac and Wolf~\cite{Banuls:2007odu} and Friis, Lee and Bruschi~\cite{Friis:2013mya} discussed how the parity superselection rule restricts the observables and the operations admissible for fermionic modes. The entanglement between fermionic modes seen by an inertial and an accelerated observer, analyzed by Alsing, Fuentes-Schuller, Mann and Tessier~\cite{Alsing:2006cj}, was examined in terms of the achievable violation of the Bell-CHSH inequality by Friis, K\"ohler, Mart\'in-Mart\'inez and Bertlmann~\cite{Friis:2011fy}. Bilinears of Majorana operators are also the fermion parities measured in topological superconductors, where unpaired Majorana modes were predicted by Kitaev~\cite{Kitaev:2000nmw}. Their realization in atomic spin chains has been reviewed by Rachel and Wiesendanger~\cite{Rachel:2024gjf}, and Liu, Qian, Cao and Liu~\cite{Liu:2025eoy} proposed a nonlocality witness for six Majorana modes built solely on parity measurements. The observables considered in the present work belong to the same class, and they lead to a construction that can be solved exactly.

In this work we study the Bell-CHSH inequality in the vacuum of the free massive Majorana field in $1+1$ dimensions with the observables of the two parties given by even bilinears of smeared fields. The one-particle vectors are generated from seeds in the modular-frequency variable $\omega$, Fourier dual to the rapidity, through the Tomita operator of the right wedge and, for the left wedge, through its adjoint multiplied by $i$. For seeds supported at positive modular frequency we obtain an exact identity showing that the overlaps between the vectors of the two parties do not depend on the modular weights $e^{\pm\pi\omega}$, which enter only through the norms of the localized vectors. Together with disjoint supports, this reduces the Wick expansion of every correlator to a single term. Choosing the four seeds as phase rotations of a single compact bump, with a translated copy supplying the second Majorana factor of each bilinear, we obtain the Bell parameter $\BCHSH=2\sqrt2\,G\,F$, where $G$ and $F$ are elementary integrals measuring how far the primary and the auxiliary bumps sit from $\omega=0$. We then show that these bilinears generate a copy of the Pauli algebra inside the even part of each wedge algebra, and we compute the full correlation matrix of the vacuum restricted to the two copies. Its third diagonal entry $T_{zz}$, which correlates the fermion parities of the primary mode on the two sides of the horizon, stays close to $-1$ over the whole range of widths we consider. The criterion of the Horodeckis~\cite{Horodecki:1995nsk} then gives $\BCHSH=2\sqrt{T_{zz}^2+G^2F^2}$, with optimal settings that are again Majorana bilinears. With these settings we \vfive{approximately halve} the deficit from $2\sqrt2$ at small width, and we extend the violation of the classical bound to spectral widths more than twice as large as with planar settings.

We also show that the Tsirelson bound is never attained in the wedge vacuum, for any choice of dichotomic observables. A quantitative version of this statement, which we derive from the modular theory of the wedge, shows that a near-maximal violation forces the modular spectral weight of the vectors obtained by acting with Alice's observables on the vacuum to concentrate at the \vfive{spectral point $\lambda=1$} of the modular operator. The point $\lambda=1$ is thus necessary as well as sufficient. Finally, we treat Gaussian seeds centred at $\omega=0$, where the overlap identity no longer applies and the internal orthogonality of each bilinear has to be imposed through a real Gram-Schmidt step. The numerical optimum we obtain approaches $2\sqrt2$ with a deficit proportional to the square of the Gaussian width. \vfive{The numerical results suggest the same structure in the narrow-width limit, with three Majorana operators per wedge and a shared factor between the two observables of each party.} Nevertheless the whole construction is confined to wedges, the one geometry in which the modular data are explicit, and the Bell correlations of regions at finite distance, which are known to decay with the mass gap and the separation~\cite{Summers:1987fn}, are not addressed here.

This work is organized as follows. In Sec.~\ref{sec:setup} we recall the one-particle structure of the Majorana field in modular frequency and the modular maps of the wedges, and we write the standard subspaces in a polar form that makes the Unruh factors $\tanh(\pi\omega)$ and $\sech(\pi\omega)$ explicit. Sec.~\ref{sec:bilinears} introduces the bilinears and their Wick structure. The overlap identity is derived in Sec.~\ref{sec:identity}, and the closed-form Bell parameter for compact seeds in Sec.~\ref{sec:compact}. In Sec.~\ref{sec:pauli} we present the local Pauli algebra, the exact correlation matrix and the optimal settings. In Sec.~\ref{sec:tsirelson} we discuss the role of the modular point and show that the Tsirelson bound is not attained. Gaussian seeds are treated in Sec.~\ref{sec:gaussian}, and in Sec.~\ref{sec:conclusions} we present our conclusions. The modular background, the proof of the bound of Sec.~\ref{sec:tsirelson} and the numerical implementation are collected in the appendices.

\section{The Majorana field in modular frequency}\label{sec:setup}

We consider the free Majorana field of mass $m$ in $1+1$-dimensional Minkowski spacetime. Its one-particle states are labelled by the rapidity $\theta$ of the on-shell momentum $p(\theta)=m(\cosh\theta,\sinh\theta)$, so that the one-particle Hilbert space is $\mathcal H_1=L^2(\mathbb R,d\theta)$ and Lorentz boosts act as translations of $\theta$. The smeared field $\psi(f)=a(f)+a^\dagger(f)$, with $a(f)$ antilinear in $f\in\mathcal H_1$, is self-adjoint and satisfies
\begin{equation}
\{\psi(f),\psi(g)\}=2\,\re\langle f,g\rangle\,\one,\qquad \langle\Omega|\psi(f)\psi(g)|\Omega\rangle=\langle f,g\rangle,
\label{eq:car}
\end{equation}
where $\langle f,g\rangle$ is the scalar product of $\mathcal H_1$, antilinear in its first argument, and $\Omega$ is the Fock vacuum. This is the quasi-free state of the self-dual CAR algebra of Araki~\cite{Araki:1971id} determined by the two-point function in \eqref{eq:car}. In particular $\psi(f)^2=\|f\|^2\one$, and fields labelled by vectors orthogonal with respect to the real part of the scalar product anticommute. We work throughout in the modular-frequency representation
\begin{equation}
f(\omega)=\frac{1}{\sqrt{2\pi}}\int_{-\infty}^{\infty}d\theta\,e^{-i\omega\theta}f(\theta),\qquad \langle f,g\rangle=\int_{-\infty}^{\infty}d\omega\,\overline{f(\omega)}\,g(\omega),
\label{eq:fourier}
\end{equation}
in which the one-particle boost generator $K=-i\,d/d\theta$ acts as multiplication by $\omega$.

By the theorem of Bisognano and Wichmann~\cite{Bisognano:1975ih,Bisognano:1976za}, the modular group of the von Neumann algebra $\mathcal M(W_R)$ of the right wedge $W_R=\{x:x^1>|x^0|\}$ in the vacuum is the group of boosts preserving $W_R$, with the rapidity rescaled by $-2\pi$. At the one-particle level this gives the modular operator $\delta=e^{-2\pi K}$ and the Tomita operator $s=j\delta^{1/2}$, where $(\delta^{1/2}f)(\theta)=f(\theta+i\pi)$ on vectors admitting an analytic continuation to the strip $0<\im\theta<\pi$ and $j$ is the one-particle modular conjugation. For the Majorana field the analytic continuation of the spinor wave functions produces a constant phase, which we absorb in a unitary redefinition of $\mathcal H_1$ that leaves the two-point function \eqref{eq:car} unchanged, so that $(jf)(\theta)=\overline{f(\theta)}$. In modular frequency the one-particle modular objects read
\begin{equation}
\begin{aligned}
(\delta f)(\omega)&=e^{-2\pi\omega}f(\omega), &\qquad (jf)(\omega)&=\overline{f(-\omega)},\\
(sf)(\omega)&=e^{\pi\omega}\,\overline{f(-\omega)}, &\qquad (s^\dagger f)(\omega)&=e^{-\pi\omega}\,\overline{f(-\omega)},
\end{aligned}
\label{eq:modmaps}
\end{equation}
and the modular spectral value $\lambda(\omega)=e^{-2\pi\omega}$ equals one at $\omega=0$. We shall refer to the \vfive{spectral point $\lambda=1$} of the modular operator, reached at $\omega=0$, as the \emph{modular point}; the distance of the spectral support of a vector from it will control all the Bell parameters computed below. Note that the mass $m$ does not appear in \eqref{eq:modmaps}. Since everything below is computed from one-particle scalar products in $\omega$, all the Bell parameters obtained in this work are independent of $m$. The same construction applies in higher dimensions: there the one-particle space factorizes as $L^2(\mathbb R,d\theta)\otimes L^2(\mathbb R^{d-2},d^{d-2}k_\perp)$, the wedge modular operator acts as $e^{-2\pi K}\otimes\one$, and for seeds of product form $\hat u(\omega)\,\chi(k_\perp)$ with $\chi$ normalized all the scalar products used below are unchanged, so that the transverse momenta enter only as spectators.

The one-particle vectors localized in $W_R$ form the standard subspace $H(W_R)=\{f\in\mathcal D(s):sf=f\}$, and $\psi(f)$ belongs to $\mathcal M(W_R)$ for every $f\in H(W_R)$. The real-linear map $1+s$ sends every $\hat u\in\mathcal D(s)$ into $H(W_R)$, since $s^2=1$ on $\mathcal D(s)$, and we call $\hat u$ the seed of the localized vector $f=(1+s)\hat u$. The symplectic complement $H(W_R)'=\{k\in\mathcal H_1:\im\langle k,f\rangle=0\ \forall f\in H(W_R)\}$ is the fixed-point space of $s^\dagger$. For the canonical anticommutation relations \eqref{eq:car} the relevant complement is the real-orthogonal one, and multiplication by $i$ turns one into the other, since $\re\langle f,ik\rangle=-\im\langle f,k\rangle$. Twisted duality, as established by Foit~\cite{Foit:1983pbi} and by Baumg\"artel, Jurke and Lled\'o~\cite{Baumgartel:1999vc}, states at the one-particle level that the left-wedge subspace is precisely this complement,
\begin{equation}
H(W_L)=i\,H(W_R)'=\{g\in\mathcal H_1:\ \re\langle g,f\rangle=0\ \ \forall f\in H(W_R)\},
\label{eq:twisted}
\end{equation}
so that the vectors of the left wedge are obtained from seeds through $g=i(1+s^\dagger)\hat v$. At the level of the algebras the same statement reads $\mathcal M(W_R)'=Z\mathcal M(W_L)Z^*$, where the twist $Z$ reduces to the identity on even operators, and this is why bilinears localized in complementary wedges commute in the ordinary sense.

A description of $H(W_R)$ that will be useful in Secs.~\ref{sec:tsirelson} and~\ref{sec:gaussian} follows from the relation $j\,\delta^{c}=\delta^{-c}j$, valid for every real $c$ as a consequence of \eqref{eq:modmaps}. Consider the real subspace
\begin{equation}
\Rr=\{r\in\mathcal H_1:\ \overline{r(-\omega)}=r(\omega)\},
\label{eq:R}
\end{equation}
formed by real even functions of $\omega$ plus $i$ times real odd ones. For $r\in\Rr$ one has $s\,\delta^{-1/4}r=j\delta^{1/4}r=\delta^{-1/4}jr=\delta^{-1/4}r$, and the converse holds as well, which gives the polar form
\begin{equation}
H(W_R)=\delta^{-1/4}\Rr,\qquad H(W_R)'=\delta^{1/4}\Rr,\qquad H(W_L)=i\,\delta^{1/4}\Rr,
\label{eq:polar}
\end{equation}
of which the maps acting on the seeds are parametrizations, $(1+s)\hat u=\delta^{-1/4}(1+j)\delta^{1/4}\hat u$. For Alice's vectors $u_k=\delta^{-1/4}r_k$ and Bob's vectors $v=i\delta^{1/4}r'$ the scalar products read
\begin{equation}
\langle u_1,u_2\rangle=\langle r_1,\cosh(\pi\omega)\,r_2\rangle+\langle r_1,\sinh(\pi\omega)\,r_2\rangle,\qquad \langle u_1,v\rangle=i\,\langle r_1,r'\rangle,
\label{eq:polar-overlaps}
\end{equation}
where the first term is real and the second purely imaginary, since $\cosh$ is even and $\sinh$ odd in $\omega$ and $\langle r,r'\rangle$ is real on $\Rr$. For a narrow band of modular frequencies around $\pm\omega_0$ this is the fermionic Unruh state, mode by mode. Alice can build two Majorana operators from the band. After normalization their bilinear has vacuum expectation value $\pm\tanh(\pi\omega_0)$, the thermal parity at inverse boost temperature $2\pi$, while the correlation of each of them with its partner on Bob's side is $\sech(\pi\omega_0)$. The identity $\tanh^2+\sech^2=1$ expresses the purity of the vacuum on the four modes. At $\omega_0=0$ each Majorana operator $\psi(\delta^{-1/4}r)$ of Alice is locally maximally mixed and perfectly paired with $\psi(i\delta^{1/4}r)$ on Bob's side, the two combining into an annihilation operator $a(r)$ with $a(r)\Omega=0$. This pairing across the horizon is the mechanism behind every construction in this work.

\section{Majorana bilinears}\label{sec:bilinears}

For two vectors $f,h\in H(W_R)$ with $\re\langle f,h\rangle=0$ the operator
\begin{equation}
A=i\,\frac{\psi(f)\psi(h)}{\|f\|\,\|h\|}
\label{eq:A}
\end{equation}
is self-adjoint and dichotomic. In fact, the two factors anticommute by \eqref{eq:car}, so that $A^\dagger=-i\,\psi(h)\psi(f)/(\|f\|\,\|h\|)$ coincides with $A$, and $A^2=\one$ because $\psi(f)^2\psi(h)^2=\|f\|^2\|h\|^2\,\one$. Bob's observables $B=i\,\psi(p)\psi(g)/\|p\|\|g\|$ are built in the same way from $g,p\in H(W_L)$ with $\re\langle g,p\rangle=0$, and by \eqref{eq:twisted} every Majorana factor of Alice anticommutes with every factor of Bob, so that $[A,B]=0$. With two such observables for each party we consider the Bell-CHSH operator and its vacuum expectation value,
\begin{equation}
\mathcal C=(A+A')B+(A-A')B',\qquad \BCHSH=\langle\Omega|\mathcal C|\Omega\rangle=\langle AB\rangle+\langle A'B\rangle+\langle AB'\rangle-\langle A'B'\rangle,
\label{eq:chsh}
\end{equation}
which is bounded by $2$ in any local hidden-variable description and by $2\sqrt2$ in quantum theory~\cite{Cirelson:1980ry}. Each correlator is a four-point function of smeared fields, and Wick's theorem for the quasi-free vacuum gives
\begin{equation}
\langle AB\rangle=-\,\frac{\langle f,h\rangle\langle p,g\rangle-\langle f,p\rangle\langle h,g\rangle+\langle f,g\rangle\langle h,p\rangle}{\|f\|\,\|h\|\,\|p\|\,\|g\|}.
\label{eq:wick}
\end{equation}
All six scalar products in \eqref{eq:wick} are purely imaginary, the internal ones because of the orthogonality conditions and the mixed ones because of \eqref{eq:twisted}. We write $\langle x,y\rangle=i\,\Gamma_{xy}\|x\|\|y\|$, where $\Gamma_{xy}=\im\langle\hat x,\hat y\rangle$ is the symplectic form evaluated on the normalized vectors $\hat x=x/\|x\|$. The correlator then becomes the Pfaffian of the antisymmetric matrix $\Gamma$ restricted to $(\hat f,\hat h,\hat p,\hat g)$,
\begin{equation}
\langle AB\rangle=\Gamma_{fh}\Gamma_{pg}-\Gamma_{fp}\Gamma_{hg}+\Gamma_{fg}\Gamma_{hp}.
\label{eq:pfaffian}
\end{equation}
Thus the correlators of Majorana bilinears are fixed by the covariance matrix of the vacuum on the real directions used by the two parties, and \eqref{eq:pfaffian} provides the natural bridge with the language of fermionic Gaussian states used in quantum information.

\section{An exact overlap identity}\label{sec:identity}

Consider two seeds $\hat u,\hat v$ and the localized vectors $f_u=(1+s)\hat u\in H(W_R)$ and $g_v=i(1+s^\dagger)\hat v\in H(W_L)$, which by \eqref{eq:modmaps} are given by
\begin{equation}
f_u(\omega)=\hat u(\omega)+e^{\pi\omega}\,\overline{\hat u(-\omega)},\qquad g_v(\omega)=i\,\big[\hat v(\omega)+e^{-\pi\omega}\,\overline{\hat v(-\omega)}\big].
\label{eq:fg}
\end{equation}
Their scalar product follows after the change of variable $\omega\to-\omega$ in the cross terms, and reads
\begin{equation}
\langle f_u,g_v\rangle=2i\,\Big[\re\langle\hat u,\hat v\rangle+\re\int_{-\infty}^{\infty}d\omega\,e^{-\pi\omega}\,\hat u(\omega)\,\hat v(-\omega)\Big],
\label{eq:overlap-general}
\end{equation}
where the modular weights have disappeared from the first term because the components of the two vectors at negative frequency carry the weights $e^{\pi\omega}$ and $e^{-\pi\omega}$, whose product is one. When both seeds are supported on the positive half-line $\omega>0$ the remaining integral vanishes identically and we obtain
\begin{equation}
\langle f_u,g_v\rangle=2i\,\re\langle\hat u,\hat v\rangle,
\label{eq:identity}
\end{equation}
exactly, for every width of the seeds and every distance of their supports from $\omega=0$. The identity \eqref{eq:identity} shows at once that the overlaps between the two parties are purely imaginary, as required by \eqref{eq:twisted}, and that they carry no information on the modular weights. These survive only in the norms: for normalized seeds supported at $\omega>0$ we have
\begin{equation}
\|f_u\|^2=1+\alpha_u,\qquad \|g_v\|^2=1+\beta_v,\qquad \alpha_u=\int d\omega\,e^{-2\pi\omega}\,|\hat u(\omega)|^2,\qquad \beta_v=\int d\omega\,e^{2\pi\omega}\,|\hat v(\omega)|^2,
\label{eq:norms}
\end{equation}
so that the normalized overlap takes the form
\begin{equation}
E(\hat u,\hat v)=\frac{\langle f_u,g_v\rangle}{i\,\|f_u\|\,\|g_v\|}=\frac{2\,\re\langle\hat u,\hat v\rangle}{\sqrt{(1+\alpha_u)(1+\beta_v)}}.
\label{eq:E}
\end{equation}
For $\hat u=\hat v$ the Cauchy-Schwarz inequality gives $\alpha_u\beta_u\ge1$, so that $(1+\alpha_u)(1+\beta_u)\ge(1+\sqrt{\alpha_u\beta_u})^2\ge4$ and $E(\hat u,\hat u)\le1$, with equality reached only in the limit of seeds shrinking to $\omega=0$. The factor $2/\sqrt{(1+\alpha_u)(1+\beta_u)}$ is thus a direct measure of the distance of the seed from the modular point $\lambda=1$, at which the localized vectors behave as if the modular structure were absent.

\section{Compact seeds and the Bell parameter}\label{sec:compact}

We now specify the seeds, starting from the smooth bump
\begin{equation}
b(x)=N_0\exp\Big(-\frac{1}{x(1-x)}\Big)\quad\text{for }0<x<1,\qquad b(x)=0\quad\text{otherwise},
\label{eq:bump}
\end{equation}
where $N_0\simeq101.5416$ normalizes $\int_0^1b(x)^2dx=1$, so that $b$ is symmetric about $x=1/2$. For a spectral width $a>0$ and a displacement $d\ge a$ we set
\begin{equation}
b_a(\omega)=a^{-1/2}\,b(\omega/a),\qquad q_{a,d}(\omega)=a^{-1/2}\,b\big((\omega-d)/a\big),
\label{eq:seeds}
\end{equation}
supported in $(0,a)$ and in $(d,d+a)$ respectively. Alice's primary vectors $f_\theta=(1+s)(e^{i\theta}b_a)$ and Bob's primary vectors $g_\varphi=i(1+s^\dagger)(e^{i\varphi}b_a)$ are phase rotations of a single bump. The auxiliary vectors $h=(1+s)q_{a,d}$ and $p=i(1+s^\dagger)q_{a,d}$ are built from the translated copy and are common to all the observables of each party,
\begin{equation}
A_\theta=i\,\frac{\psi(f_\theta)\psi(h)}{\|f_\theta\|\,\|h\|},\qquad B_\varphi=i\,\frac{\psi(p)\psi(g_\varphi)}{\|p\|\,\|g_\varphi\|}.
\label{eq:AthetaBphi}
\end{equation}
The primary vectors are supported in $(-a,a)$ and the auxiliary ones in $(-d-a,-d)\cup(d,d+a)$, and the disjointness of these supports has two consequences. The internal overlaps $\langle f_\theta,h\rangle$ and $\langle p,g_\varphi\rangle$ vanish identically, so that dichotomy holds by construction without any orthogonalization, and so do the mixed overlaps $\langle f_\theta,p\rangle$ and $\langle h,g_\varphi\rangle$ between the primary vector of one party and the auxiliary vector of the other. Only the last term of \eqref{eq:wick} survives, and the identity \eqref{eq:identity} gives
\begin{equation}
\langle A_\theta B_\varphi\rangle=-\,\frac{\langle f_\theta,g_\varphi\rangle\langle h,p\rangle}{\|f_\theta\|\,\|g_\varphi\|\,\|h\|\,\|p\|}=G(a)\,F(a,d)\,\cos(\theta-\varphi),
\label{eq:corr-compact}
\end{equation}
where we used $\re\langle e^{i\theta}b_a,e^{i\varphi}b_a\rangle=\cos(\theta-\varphi)$ and $\langle h,p\rangle=2i$. Here $G=2/\sqrt{(1+\alpha)(1+\beta)}$ is the suppression factor \eqref{eq:E} evaluated on $b_a$, and $F=2/\sqrt{(1+\alpha_q)(1+\beta_q)}$ the same factor evaluated on $q_{a,d}$. Using the symmetry of $b$ about $x=1/2$ we obtain $\alpha=e^{-\pi a}I(a)$, $\beta=e^{\pi a}I(a)$, $\alpha_q=e^{-\pi(2d+a)}I(a)$ and $\beta_q=e^{\pi(2d+a)}I(a)$, and therefore
\begin{equation}
G(a)=\frac{2}{\sqrt{1+2I(a)\cosh(\pi a)+I(a)^2}},\qquad F(a,d)=\frac{2}{\sqrt{1+2I(a)\cosh\big(\pi(2d+a)\big)+I(a)^2}},
\label{eq:GF}
\end{equation}
where
\begin{equation}
I(a)=\int_0^1dx\,\cosh\big(2\pi a(x-\tfrac12)\big)\,b(x)^2\ \ge\ 1 .
\label{eq:I}
\end{equation}
The factor $F$ is common to the four correlators, and the whole dependence on the settings sits in the cosine of \eqref{eq:corr-compact}, which is the correlation of unit vectors in a plane. The standard choice $\theta_1=0$, $\theta_2=\pi/2$, $\varphi_1=\pi/4$, $\varphi_2=-\pi/4$ gives
\begin{equation}
\BCHSH^{\rm planar}(a,d)=2\sqrt2\,G(a)\,F(a,d),
\label{eq:Bplanar}
\end{equation}
strictly below $2\sqrt2$ for every $a>0$. For small width, with $d=\rho a$ and $\mu_2=\int_0^1(x-\tfrac12)^2\,b(x)^2dx\simeq0.0116856$, the expansion reads
\begin{equation}
\BCHSH^{\rm planar}(a,\rho a)=2\sqrt2\,\Big[1-\pi^2a^2\Big(2\mu_2+\frac18+\frac{(2\rho+1)^2}{8}\Big)+O(a^4)\Big],
\label{eq:Bplanar-exp}
\end{equation}
which contains only even powers of $a$. At the contact configuration $\rho=1$ the coefficient is $\pi^2(2\mu_2+5/4)\simeq12.568$. Most of the deficit comes from the auxiliary term $(2\rho+1)^2/8$, because the translated bump sits further from $\omega=0$ than the primary one, and the classical bound is violated for $a<0.17361$ (Table~\ref{tab:compact}).

\section{The local Pauli algebra}\label{sec:pauli}

The planar settings of \eqref{eq:Bplanar} leave unused part of the algebra generated by the bilinears \eqref{eq:AthetaBphi}. Since $f_\theta=\cos\theta\,f_0+\sin\theta\,f_{\pi/2}$, with a norm independent of $\theta$, we have $A_\theta=\cos\theta\,A_0+\sin\theta\,A_{\pi/2}$. The three unit vectors $\hat f_0$, $\hat f_{\pi/2}$ and $\hat h$ are mutually orthogonal with respect to the real part of the scalar product, the first two because $\langle f_0,f_{\pi/2}\rangle=i(1-\alpha)$. The operators
\begin{equation}
X_A=A_0,\qquad Y_A=A_{\pi/2},\qquad Z_A=-i\,X_AY_A=-i\,\psi(\hat f_0)\psi(\hat f_{\pi/2}),
\label{eq:pauliA}
\end{equation}
therefore satisfy the relations of the Pauli matrices and generate a copy of $M_2(\mathbb C)$ inside the even part of $\mathcal M(W_R)$. The same holds on Bob's side with $X_B=B_0$, $Y_B=B_{\pi/2}$ and $Z_B=-i\,X_BY_B=-i\,\psi(\hat g_0)\psi(\hat g_{\pi/2})$. Note that $Z_A$ and $Z_B$ are the fermion parities of the primary mode on each side of the horizon. Every combination $n_1X_A+n_2Y_A+n_3Z_A$ with $n_1^2+n_2^2+n_3^2=1$, and likewise on Bob's side, is again a Majorana bilinear $i\psi(u)\psi(w)$, with $u$ and $w$ orthonormal in the real span of the three vectors of the party, because every two-form on a three-dimensional real space is decomposable. The vacuum restricted to the algebra generated by the two copies is a two-qubit density matrix, fixed by the local expectation values and by the correlation matrix $T_{jk}=\langle\Omega|\sigma^A_j\sigma^B_k|\Omega\rangle$, where $\sigma^A=(X_A,Y_A,Z_A)$ and $\sigma^B=(X_B,Y_B,Z_B)$. We compute all of them from \eqref{eq:wick}, \eqref{eq:identity} and \eqref{eq:norms}. The local terms are $\langle Z_A\rangle=(1-\alpha)/(1+\alpha)$ and $\langle Z_B\rangle=(1-\beta)/(1+\beta)$, the $x$ and $y$ components vanishing. Disjointness sets to zero every entry of $T$ that mixes $Z$ with $X$ or $Y$, leaving
\begin{equation}
T=\mathrm{diag}\big(G F,\ G F,\ T_{zz}\big),\qquad T_{zz}=\frac{(1-\alpha)(1-\beta)-4}{(1+\alpha)(1+\beta)}=-1+\frac12\,\big[I(a)^2-1\big]\,G(a)^2 ,
\label{eq:T}
\end{equation}
where the second form of $T_{zz}$ follows from $\alpha\beta=I(a)^2$ and $(1+\alpha)(1+\beta)=4/G(a)^2$.

The entry $T_{zz}$ correlates the parities of the primary mode on the two sides of the horizon, and its value close to $-1$ has a simple origin. For a sharply defined modular frequency the primary vectors of the two parties span a two-dimensional complex subspace of $\mathcal H_1$, on which the vacuum is a pure state of definite total parity, so that $T_{zz}=-1$ exactly; this is the case $I=1$ of \eqref{eq:T}. The finite width of the bump makes $I(a)>1$, through the Cauchy-Schwarz inequality $\alpha\beta\ge1$. But the correction carries only the small moment $\mu_2$, $T_{zz}=-1+2\pi^2\mu_2a^2+O(a^4)$, while the in-plane entries $GF$ decay with the distance of both bumps from $\omega=0$. The criterion of the Horodeckis~\cite{Horodecki:1995nsk} gives the maximum of the Bell parameter over all dichotomic observables of the two copies as $2\sqrt{t_1^2+t_2^2}$, where $t_1$ and $t_2$ are the two largest singular values of $T$. In the whole range of widths considered here $|T_{zz}|>GF$, and we obtain
\begin{equation}
\BCHSH^{\rm opt}(a,d)=2\sqrt{T_{zz}^2+G(a)^2F(a,d)^2}\ \ \ge\ \ 2\sqrt2\,G(a)\,F(a,d).
\label{eq:Bopt}
\end{equation}
The maximum is reached with Alice measuring $A=Z_A$ and $A'=X_A$ and Bob measuring $B=-\cos\gamma\,Z_B+\sin\gamma\,X_B$ and $B'=-\cos\gamma\,Z_B-\sin\gamma\,X_B$, with $\tan\gamma=GF/|T_{zz}|$. For these settings $\langle AB\rangle=\langle AB'\rangle=|T_{zz}|\cos\gamma$ and $\langle A'B\rangle=-\langle A'B'\rangle=GF\sin\gamma$. All four observables are Majorana bilinears, $A=-i\psi(\hat f_0)\psi(\hat f_{\pi/2})$, $A'=i\psi(\hat f_0)\psi(\hat h)$ and
\begin{equation}
B=i\,\psi(\hat g_0)\psi(w_-),\qquad B'=i\,\psi(\hat g_0)\psi(w_+),\qquad w_\pm=\cos\gamma\,\hat g_{\pi/2}\pm\sin\gamma\,\hat p,
\label{eq:optsettings}
\end{equation}
so that the Majorana factor shared by the two observables of each party is a primary vector, whereas in the planar settings it is the auxiliary one. One of Alice's observables is the parity of a single modular mode, which correlates almost perfectly with Bob's, and the remaining settings only have to extract the in-plane correlation $GF$.

\begin{figure}[tbp]
\centering
\includegraphics[width=0.62\textwidth]{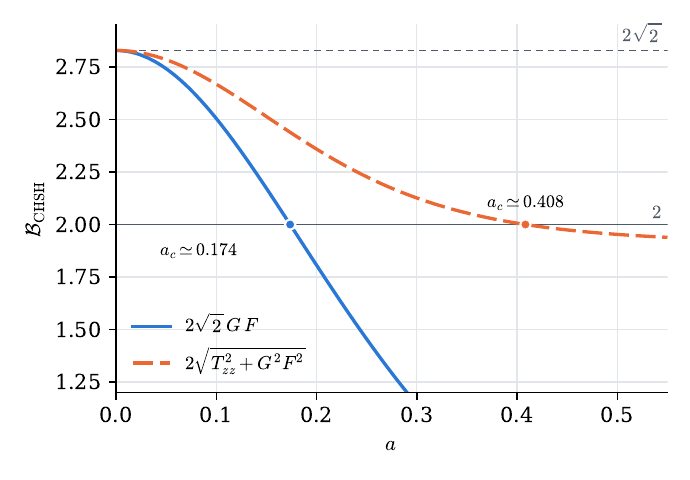}
\caption{Bell parameter for compact seeds at contact, $d=a$, as a function of the spectral width $a$, with the planar settings \eqref{eq:Bplanar} (solid) and with the optimal settings \eqref{eq:Bopt} in the local Pauli algebra (dashed). The dots mark the widths at which each curve crosses the classical bound; the dashed horizontal line is the Tsirelson value.}
\label{fig:compact}
\end{figure}

For small width \eqref{eq:Bopt} expands as
\begin{equation}
\BCHSH^{\rm opt}(a,\rho a)=2\sqrt2\,\Big[1-\pi^2a^2\Big(2\mu_2+\frac{1+(2\rho+1)^2}{16}\Big)+O(a^4)\Big],
\label{eq:Bopt-exp}
\end{equation}
to be compared with \eqref{eq:Bplanar-exp}: the terms coming from the positions of the two bumps are halved, while the shape term $2\mu_2$ is unchanged. At contact the deficit is $2\sqrt2\,\pi^2(2\mu_2+5/8)\,a^2\simeq18.10\,a^2$, against $35.55\,a^2$ for the planar settings, and the classical bound is now violated for $a<0.40829$, more than twice the planar value. Table~\ref{tab:compact} and Fig.~\ref{fig:compact} compare the two sets of settings, and the improvement persists away from contact: for $a=0.04$ and $d=2a$ the planar settings give $2.6881$ and the optimal ones $2.7586$.

\begin{table}[tbp]
\centering
\begin{tabular}{cccccc}
\toprule
$a$ & $G(a)$ & $F(a,a)$ & $T_{zz}$ & $\BCHSH^{\rm planar}$ & $\BCHSH^{\rm opt}$\\
\midrule
0.001 & 0.999999 & 0.999989 & $-1.000000$ & 2.82839 & 2.82841\\
0.01 & 0.999865 & 0.998879 & $-0.999977$ & 2.82488 & 2.82662\\
0.04 & 0.997845 & 0.982313 & $-0.999632$ & 2.77241 & 2.80004\\
0.10 & 0.986650 & 0.897354 & $-0.997750$ & 2.50421 & 2.66788\\
0.15 & 0.970346 & 0.791341 & $-0.995090$ & 2.17188 & 2.51383\\
0.20 & 0.948212 & 0.673476 & $-0.991633$ & 1.80623 & 2.35893\\
0.30 & 0.889087 & 0.454559 & $-0.983268$ & 1.14309 & 2.12617\\
0.40 & 0.815340 & 0.291410 & $-0.974603$ & 0.67203 & 2.00629\\
\bottomrule
\end{tabular}
\caption{Compact seeds at contact, $d=a$: suppression factors \eqref{eq:GF}, parity correlation $T_{zz}$ of \eqref{eq:T}, and the Bell parameter with the planar settings \eqref{eq:Bplanar} and with the optimal settings \eqref{eq:Bopt}. For comparison, $2\sqrt2=2.828427\ldots$}
\label{tab:compact}
\end{table}

Narrow bumps centred at fixed modular frequencies give a useful limiting form of \eqref{eq:Bopt}. If the primary seed is concentrated near $\omega_1>0$ and the auxiliary one near $\omega_2>0$, then $\alpha\to e^{-2\pi\omega_1}$, $\beta\to e^{2\pi\omega_1}$ and $I\to1$, so that $G\to\sech(\pi\omega_1)$, $F\to\sech(\pi\omega_2)$ and $T_{zz}\to-1$, and
\begin{equation}
\BCHSH^{\rm opt}\ \longrightarrow\ 2\sqrt{1+\sech^2(\pi\omega_1)\,\sech^2(\pi\omega_2)}\ \simeq\ 2\sqrt2-\frac{\pi^2}{\sqrt2}\,\big(\omega_1^2+\omega_2^2\big),
\label{eq:Bsharp}
\end{equation}
which exceeds the classical bound for all frequencies and reaches $2\sqrt2$ only when both frequencies go to zero. The factors $\sech(\pi\omega_k)$ are the Unruh coherences of Sec.~\ref{sec:setup}, and \eqref{eq:Bsharp} shows that, within the algebra generated by three Majorana operators per party, the bilinears exploit them optimally once the parity correlation of a single mode is used as one of the two measurement axes. We have checked \eqref{eq:Bsharp} against narrow Gaussian packets and against an unrestricted optimization over all bilinears built from the three Majorana operators of each party, which returns the same value. With the four Majorana operators provided by two modes on each side the unrestricted optimum over bilinears is slightly higher, for instance $2.333$ instead of $2.294$ at $(\omega_1,\omega_2)=(0.2,0.3)$.

\section{The modular point and the Tsirelson bound}\label{sec:tsirelson}

All the families considered so far approach $2\sqrt2$ only in a limit, and this is not a feature of the particular seeds. Consider arbitrary dichotomic observables $A,A'\in\mathcal M(W_R)$ and $B,B'\in\mathcal M(W_R)'$. Since they square to one and commute across the two parties, the Bell-CHSH operator admits the sum-of-squares decomposition
\begin{equation}
2\sqrt2\,\one-\mathcal C=\frac{1}{\sqrt2}\Big[(A-y)^2+(A'-y')^2\Big],\qquad y=\frac{B+B'}{\sqrt2},\qquad y'=\frac{B-B'}{\sqrt2},
\label{eq:sos}
\end{equation}
which reproduces the bound of Tsirelson and shows that the value $2\sqrt2$ can be reached in the vacuum only if $A\Omega=y\,\Omega$ and $A'\Omega=y'\,\Omega$. This is the rigidity of maximal violation discussed by Popescu and Rohrlich~\cite{Popescu:1992yq}. If this happens, the Tomita operator $S$ of $(\mathcal M(W_R),\Omega)$ gives $SA\Omega=A\Omega$, because $A$ is self-adjoint and belongs to $\mathcal M(W_R)$. In the same way its adjoint $S^*$, which is the Tomita operator of the commutant, gives $S^*y\Omega=y\Omega$, because $y$ is self-adjoint and belongs to $\mathcal M(W_R)'$. The vector $\xi=A\Omega=y\Omega$ then satisfies both $S\xi=\xi$ and $S^*\xi=\xi$, that is $\Delta^{1/2}\xi=J\xi$ and $\Delta^{-1/2}\xi=J\xi$, where $S=J\Delta^{1/2}$, $S^*=J\Delta^{-1/2}$, and we used that $J$ is an involution. Applying $J$ to the second relation and using $J\Delta^{-1/2}J=\Delta^{1/2}$ gives $\Delta^{1/2}(J\xi)=\xi$, so that $\Delta^{1/2}\xi=J\xi$ lies in the domain of $\Delta^{1/2}$. Therefore $\xi\in D(\Delta)$ and $\Delta\xi=\Delta^{1/2}\big(\Delta^{1/2}\xi\big)=\Delta^{1/2}(J\xi)=\xi$, and by the theorem of Bisognano and Wichmann $\xi$ is invariant under all the boosts preserving the wedge. For the free Majorana field the boost generator restricted to the $n$-particle space acts as multiplication by the total modular frequency $\omega_1+\dots+\omega_n$, whose spectrum is purely absolutely continuous for $n\ge1$, so that the only boost-invariant vectors are the multiples of $\Omega$. Thus $A\Omega=c\,\Omega$, and since $\Omega$ is separating for $\mathcal M(W_R)$ we conclude that $A=c\,\one$ with $c=\pm1$. The same argument applies to $A'$, and with $A$ and $A'$ both multiples of the identity the Bell parameter cannot exceed $2$, which is a contradiction. We have therefore shown that the Tsirelson bound is never attained in the wedge vacuum of the free Majorana field, for any dichotomic observables of $\mathcal M(W_R)$ and of its commutant, bilinear or not. The argument uses only the Bisognano-Wichmann property and the absence of boost-invariant vectors other than the vacuum. It applies without change to the scalar field and to the Weyl operators of~\cite{Caribe:2026mam}, and to the half-line algebras of a chiral theory with dilations in place of boosts. \vfive{It also extends to Poincar\'e-covariant theories satisfying the Bisognano-Wichmann property, the spectrum condition and vacuum uniqueness.} \vfive{By the standard argument of Mautner type}, a vector invariant under the boosts is invariant under the lightlike translations, which the boosts contract to the identity, and the spectrum condition together with the uniqueness of the vacuum then makes it a multiple of $\Omega$. In all these cases $2\sqrt2$ is a supremum that no configuration of observables attains, and this is the sense of the maximal violation of Summers and Werner~\cite{Summers:1987squ,Summers:1987ze} and of the results of~\cite{Caribe:2026mam,Caribe:2026xoa,Caribe:2026tfe}. \vfive{The qualitative non-attainment statement also follows from the centralizer criterion of Summers and Werner~\cite{Summers:1987fn}. Exact saturation requires the observables to be fixed by the modular flow. In the wedge vacuum considered here, the separating property and the absence of boost-invariant vectors orthogonal to the vacuum make this fixed-point algebra trivial. We give a direct modular proof and obtain the quantitative estimate~\eqref{eq:necessity}, which controls the modular spectral concentration required for near-maximal violation.}

The same reasoning gives a quantitative statement for a near-maximal violation. If $\BCHSH=2\sqrt2-\varepsilon$ with $\varepsilon<1/\sqrt2$, then
\begin{equation}
\big\|(\Delta^{1/2}-1)\,A\Omega\big\|^2\ \le\ \frac{2\sqrt2\,\varepsilon}{1-2^{1/4}\sqrt{\varepsilon}},
\label{eq:necessity}
\end{equation}
and the same bound holds for $A'\Omega$, and for $B\Omega$ and $B'\Omega$ with $\Delta^{-1/2}$ in place of $\Delta^{1/2}$, as we show in Appendix~\ref{app:necessity}. The left-hand side of \eqref{eq:necessity} is the integral of $(\sqrt\lambda-1)^2$ over the spectral measure of $\Delta$ in the vector $A\Omega$. The bound therefore states that the modular spectral weight of the vectors produced by Alice's observables must concentrate at $\lambda=1$ as the violation approaches its supremum. The point $\lambda=1$ singled out in the Summers-Werner mechanism is thus necessary, while the constructions of Secs.~\ref{sec:compact} and~\ref{sec:pauli} show that it is also sufficient. For a bilinear $A=i\psi(\hat f)\psi(\hat h)$ the vector $(\Delta^{1/2}-1)A\Omega$ lies in the two-particle space, where $\Delta^{1/2}$ acts as multiplication by $e^{-\pi(\omega_1+\omega_2)}$. The bound \eqref{eq:necessity} then constrains the distribution of the total modular frequency of the pair $\hat f\wedge\hat h$, with $\pi^2\langle(\omega_1+\omega_2)^2\rangle\lesssim2\sqrt2\,\varepsilon$ at leading order. One can read the structure of the optimal settings \eqref{eq:optsettings} in these terms. The parity $Z_A$ of a single mode pairs its components at $\pm\omega_1$ and has vanishing total modular frequency in the limit of a sharp mode, so that it costs nothing in \eqref{eq:necessity}. The operator $X_A$ couples the primary and the auxiliary bumps instead, and pays for both of their frequencies. With the settings \eqref{eq:optsettings} and the compact seeds of Sec.~\ref{sec:compact} we have checked numerically that \eqref{eq:necessity} holds with a margin of about a factor of two; at $a=d=0.005$, for instance, $\|(\Delta^{1/2}-1)A'\Omega\|^2=6.2\times10^{-4}$ against a bound of $1.3\times10^{-3}$.

\section{Gaussian seeds centred at the modular point}\label{sec:gaussian}

The compact seeds of Sec.~\ref{sec:compact} are supported away from $\omega=0$. We now consider seeds centred at the modular point itself, as in the Hermite-Gaussian construction from which the present work started. Two things change: the second term of \eqref{eq:overlap-general} no longer vanishes, and the internal orthogonality of each bilinear has to be imposed explicitly. Since the vectors of each party form a real subspace, this is achieved by the real Gram-Schmidt step
\begin{equation}
v_\perp=v-\frac{\re\langle u,v\rangle}{\|u\|^2}\,u,
\label{eq:gs}
\end{equation}
which involves the real part of the scalar product, that is the metric entering \eqref{eq:car}. The step preserves localization, since $v_\perp$ is a real combination of vectors of the same subspace. Note also that for a real seed $\varphi$ the vectors $(1+s)\varphi$ and $(1+s)\,i\varphi$ are automatically real-orthogonal, because $\re\langle(1+s)\varphi,(1+s)i\varphi\rangle=-2\,\im\langle s\varphi,\varphi\rangle$ and $\langle s\varphi,\varphi\rangle=\int d\omega\,e^{\pi\omega}\varphi(\omega)\varphi(-\omega)$ is real.

As trial space we take the Hermite-Gaussian functions of width $\sigma$,
\begin{equation}
\chi_n^{(\sigma)}(\omega)=e^{-\omega^2/2\sigma^2}\,\frac{H_n(\omega/\sigma)}{\sqrt{2^n\,n!}},\qquad n=0,\dots,N-1,
\label{eq:hermite}
\end{equation}
where $H_n$ are the Hermite polynomials. We parametrize the vectors of both parties through the polar form \eqref{eq:polar}, writing $u=\delta^{-1/4}r$ for Alice and $v=i\delta^{1/4}r'$ for Bob, with $r,r'$ in the real span $\Rr_N$ of $\chi_0,i\chi_1,\chi_2,i\chi_3,\dots$, which is the intersection of \eqref{eq:R} with the trial space. \vfive{The numerical results reported below refer to this polar trial space.} Each party uses four independent vectors, orthogonalized in pairs by \eqref{eq:gs}. We evaluate the correlators from \eqref{eq:wick} with Gauss-Hermite quadrature and maximize the Bell parameter with a quasi-Newton method started from many random points, as described in Appendix~\ref{app:numerics}. The values we report are therefore lower bounds on the supremum over each trial space.

\begin{figure}[tbp]
\centering
\includegraphics[width=0.95\textwidth]{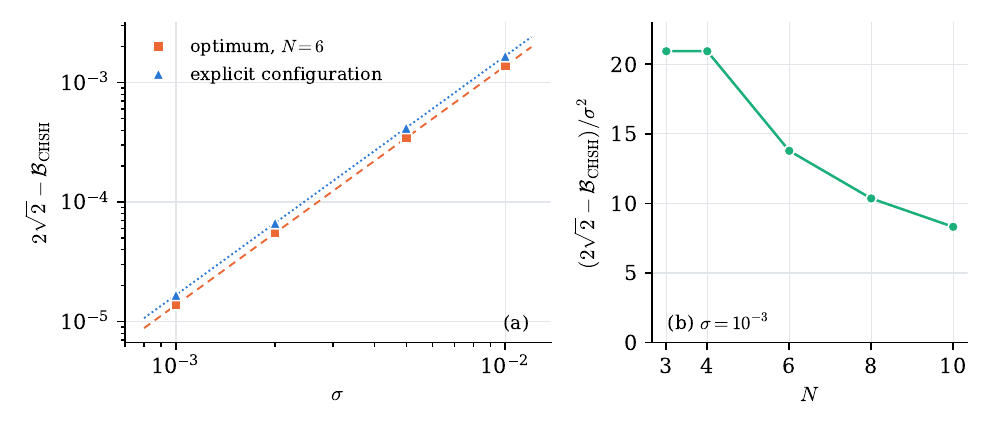}
\caption{Gaussian seeds centred at $\omega=0$. (a) Deficit $2\sqrt2-\BCHSH$ as a function of the width $\sigma$ with $N=6$ Hermite functions: numerical optimum (squares) and explicit configuration without optimization (triangles); the lines are $13.78\,\sigma^2$ and $16.71\,\sigma^2$. (b) Coefficient of $\sigma^2$ in the deficit at $\sigma=10^{-3}$ as a function of the number $N$ of Hermite functions in the trial space.}
\label{fig:gaussian}
\end{figure}

With $N=6$ modes the optimum approaches the Tsirelson bound with a deficit $2\sqrt2-\BCHSH\simeq13.78\,\sigma^2$, with a coefficient constant to within $0.1\%$ over the whole range $10^{-3}\le\sigma\le10^{-2}$ [Fig.~\ref{fig:gaussian}(a)]. At $\sigma=10^{-3}$ this gives $\BCHSH=2.8284133$. An explicit configuration without any optimization already reaches $16.71\,\sigma^2$. In it Alice's three Majorana vectors are $\delta^{-1/4}r_k$, with $r_1$ and $r_3$ the two lowest eigenvectors of the compression of $\omega^2$ to the even part of $\Rr_6$ and $r_2$ the lowest eigenvector of its compression to the odd part. Bob uses $i\delta^{1/4}r_k$ with the same $r_k$, and we take the settings from the criterion of the Horodeckis as in Sec.~\ref{sec:pauli}. The coefficient of $\sigma^2$ is a property of the trial space and decreases as modes are added: at $\sigma=10^{-3}$ it is $20.9$ for $N=3$ and $N=4$, $13.8$ for $N=6$, $10.4$ for $N=8$ and $8.3$ for $N=10$ [Fig.~\ref{fig:gaussian}(b)]. This indicates that the approach to $2\sqrt2$ is controlled by the effective spectral width of the vectors in $\omega$, which can be reduced either by lowering $\sigma$ or by admitting narrower combinations of Hermite functions, in agreement with \eqref{eq:necessity}.

\vfive{The optimized configurations exhibit approximately the structure found for the compact seeds. The real Gram matrices of the four normalized vectors on each side have eigenvalues close to $\{0,1,1,2\}$. These numerical results suggest a three-direction structure in the narrow-width limit, with a shared Majorana factor and asymptotically anticommuting bilinears, as in Sec.~\ref{sec:pauli}.} The four correlators approach the pattern $(1/\sqrt2,1/\sqrt2,1/\sqrt2,-1/\sqrt2)$ individually, as they must by the rigidity expressed by \eqref{eq:sos}. At $\sigma=10^{-3}$ with $N=6$ they are $0.7071088$, $0.7071016$, $0.7071051$ and $-0.7070978$.

\section{Conclusions}\label{sec:conclusions}

In this work we have obtained in closed form the Bell-CHSH parameter of the Majorana vacuum in complementary wedges, for dichotomic observables given by even bilinears of smeared fields generated from compact seeds in modular frequency. The construction started from the observation that for seeds supported at positive modular frequency the overlaps between the two parties are independent of the modular weights, which reduces every correlator to a single Wick term. This led us to recognize that the bilinears generate a copy of the Pauli algebra in each wedge. The vacuum restricted to the two copies is a two-qubit state with correlation matrix \eqref{eq:T}. The optimal settings \eqref{eq:optsettings}, which are again bilinears, give $\BCHSH=2\sqrt{T_{zz}^2+G^2F^2}$. They \vfive{approximately halve} the deficit from $2\sqrt2$ at small width and extend the violation of the classical bound to spectral widths more than twice as large as those allowed by planar settings. We have also shown that the Tsirelson bound is never attained in the wedge vacuum, for any dichotomic observables. The bound we have derived for a near-maximal violation forces the modular spectral weight of the Bell vectors to concentrate at the \vfive{spectral point $\lambda=1$} of the modular operator, so that the point singled out by Summers and Werner is necessary as well as sufficient. Gaussian seeds centred at $\omega=0$, treated numerically, approach $2\sqrt2$ with a deficit proportional to $\sigma^2$, whose coefficient decreases with the number of modes of the trial space, and \vfive{suggest the same architecture in the narrow-width limit}.

Nevertheless these results are confined to wedges, where the Bisognano-Wichmann theorem gives complete control of the modular data. The Bell correlations of regions at finite distance, which decay with the mass and the separation~\cite{Summers:1987fn}, remain outside the present analysis. \vfive{For the massless Dirac field in $1+1$ dimensions, Casini and Huerta~\cite{Casini:2009vk} have obtained explicit modular dynamics for unions of disjoint intervals.} Bilinears seem well suited to a computation of the Bell parameter as a function of the cross ratio of the two intervals, which we leave for future work. On the side of condensed matter, the bilinears used here are fermion parities, and the minimal structure we obtain, three Majorana operators per party with one shared factor, coincides with the one probed by the parity-based witness of~\cite{Liu:2025eoy}. It is important to stress that no laboratory system realizes the modular flow of a Rindler wedge. What can be carried over is the algebraic skeleton of the construction, the dichotomic parities and their pattern of commutation, while the approach to $\lambda=1$ that produces the near-maximal violation has at present no laboratory counterpart. This provides an explicit characterization of how the even sector of the free Majorana field approaches the Tsirelson bound: through three Majorana operators per wedge, one of whose bilinears is the parity of a single modular mode, with a deficit fixed by the distance of the seeds from the \vfive{spectral point $\lambda=1$} of the modular operator.

\section*{Acknowledgments}
% TODO(authors): confirm the grant numbers (S.P.S. taken from note #08; I.R. and M.S.G. from the
% Bell-thermal manuscript), and confirm the AI-use statement against the
% policy of the target journal.
The authors thank the Brazilian agencies CNPq, CAPES and FAPERJ for financial support. S.~P.~Sorella, I.~Roditi and M.~S.~Guimaraes are CNPq researchers under contracts 302991/2024-7, 319060/2025-0 and 309793/2023-8, respectively. The authors acknowledge the use of AI assistants (ChatGPT, OpenAI, and Claude, Anthropic) for exploratory calculations, numerical cross-checks and language editing; all scientific content, calculations, interpretations and conclusions were checked by the authors, who remain solely responsible for them.

\section*{Data availability}
The data and code supporting the numerical analysis of this work are openly available in the repository \url{https://github.com/silviosorella-web/Majorana-Bilinear}.

\appendix

\section{Modular background}\label{app:modular}

A closed real subspace $H$ of a complex Hilbert space $\mathcal H$ is standard if $H\cap iH=\{0\}$ and $H+iH$ is dense. Every vector of $H+iH$ then decomposes uniquely as $h_1+ih_2$ with $h_1,h_2\in H$, and the closed antilinear involution $s_H(h_1+ih_2)=h_1-ih_2$ has polar decomposition $s_H=j_H\delta_H^{1/2}$, with $j_H$ antiunitary and $\delta_H$ positive and self-adjoint. The two factors satisfy $j_H\delta_Hj_H=\delta_H^{-1}$ and $\delta_H^{it}H=H$. The symplectic complement $H'$ is standard as well, with $s_{H'}=s_H^\dagger$. For every $\phi$ in the domain of $s_H$ one has $(1+s_H)\phi\in H$, so that $1+s_H$ maps the domain onto $H$ and acts as $2$ on $H$ itself; these real-linear maps are not orthogonal projections. For $H=H(W_R)$ the Bisognano-Wichmann theorem identifies $\delta_H^{it}$ with the boosts preserving $W_R$, $\delta_H=e^{-2\pi K}$, and the continuation of the one-particle wave functions to the strip gives $(\delta^{1/2}f)(\theta)=f(\theta+i\pi)$. With the Fourier transform \eqref{eq:fourier} one has $f(\theta+i\pi)=(2\pi)^{-1/2}\int d\omega\,e^{i\omega\theta}e^{-\pi\omega}f(\omega)$, and $\overline{f(\theta)}=(2\pi)^{-1/2}\int d\omega\,e^{i\omega\theta}\,\overline{f(-\omega)}$, which give \eqref{eq:modmaps}.

The second quantization of $H(W_R)$ through the self-dual CAR algebra of Araki~\cite{Araki:1971id} produces the wedge algebra $\mathcal M(W_R)=\{\psi(f):f\in H(W_R)\}''$, for which the vacuum is cyclic and separating. The fermion parity $\Pi=(-1)^{\mathcal N}$, with $\mathcal N$ the number operator, defines the twist $Z=(1+i\Pi)/(1+i)$, which acts as $Z\psi(g)Z^*=i\Pi\psi(g)$ on odd operators and as the identity on even ones. Twisted duality~\cite{Foit:1983pbi,Baumgartel:1999vc} states that $\mathcal M(W_R)'=Z\mathcal M(W_L)Z^*$, whose one-particle content is \eqref{eq:twisted}. The polar form \eqref{eq:polar} follows from $j\delta^cj=\delta^{-c}$, a consequence of \eqref{eq:modmaps}: for $u=\delta^{-1/4}r$ one has $su=j\delta^{1/4}r=\delta^{-1/4}jr$, so that $su=u$ if and only if $jr=r$, and the same argument with $s^\dagger=\delta^{1/2}j$ gives $H(W_R)'=\delta^{1/4}\Rr$. For regions other than wedges one defines $H(\mathcal O)=\bigcap_{W\supset\mathcal O}H(W)$, the intersection running over the wedges containing $\mathcal O$; this subspace is standard for double cones in free massive theories (see Brunetti, Guido and Longo~\cite{Brunetti:2002nt}), and it is the starting point for an extension of the present construction to bounded regions.

\section{Proof of the bound on the modular spectral weight}\label{app:necessity}

Let $\BCHSH=2\sqrt2-\varepsilon$. By \eqref{eq:sos} we have $\varepsilon=\frac{1}{\sqrt2}\big(\|(A-y)\Omega\|^2+\|(A'-y')\Omega\|^2\big)$, so that $\kappa^2:=\|(A-y)\Omega\|^2\le\sqrt2\,\varepsilon$. Write $\xi=A\Omega$, of unit norm since $A$ is unitary, and $\eta=y\Omega$, whose norm $Y$ satisfies $1-\kappa\le Y\le1+\kappa$. The vector $\xi$ belongs to the domain of $\Delta^{1/2}$ with $\Delta^{1/2}\xi=J\xi$, since $SA\Omega=A\Omega$, and $\eta$ belongs to the domain of $\Delta^{-1/2}$ with $\Delta^{-1/2}\eta=J\eta$, since $S^*y\Omega=y\Omega$ and $S^*=J\Delta^{-1/2}$. Consider $\xi_+=\Delta^{1/4}\xi$ and $\eta_-=\Delta^{-1/4}\eta$. Their scalar product is $\langle\xi_+,\eta_-\rangle=\langle\xi,\eta\rangle$, while $\|\xi_+\|^2=\langle\xi,J\xi\rangle\le1$ and $\|\eta_-\|^2=\langle\eta,J\eta\rangle\le Y^2$. The Cauchy-Schwarz inequality then gives
\begin{equation}
\|\xi_+\|\ \ge\ \frac{\re\langle\xi,\eta\rangle}{Y}=\frac{1+Y^2-\kappa^2}{2Y}\ \ge\ 1-\frac{\kappa^2}{2(1-\kappa)},
\label{eq:app-cs}
\end{equation}
where we used $(1+Y^2)/2Y\ge1$. On the other hand $\|\Delta^{1/2}\xi\|=\|J\xi\|=1=\|\xi\|$, and therefore
\begin{equation}
\big\|(\Delta^{1/2}-1)\xi\big\|^2=\|\Delta^{1/2}\xi\|^2-2\,\langle\xi,\Delta^{1/2}\xi\rangle+\|\xi\|^2=2\big(1-\|\xi_+\|^2\big)\vfive{.}
\label{eq:app-norm}
\end{equation}
\vfive{Since $\kappa<1$, the arithmetic--geometric mean inequality gives $(1+Y^2-\kappa^2)/(2Y)\ge\sqrt{1-\kappa^2}$. Equations~\eqref{eq:app-cs} and~\eqref{eq:app-norm} therefore yield $\|(\Delta^{1/2}-1)\xi\|^2\le2\kappa^2\le2\sqrt2\,\varepsilon$, which implies~\eqref{eq:necessity}.} The same steps with the roles of $\mathcal M(W_R)$ and of its commutant exchanged give the bound for $B\Omega$ and $B'\Omega$ with $\Delta^{-1/2}$.

\section{Numerical implementation}\label{app:numerics}

For the Gaussian seeds of Sec.~\ref{sec:gaussian} we use the variable $x=\omega/\sigma$ and store each one-particle vector through its values, multiplied by $e^{x^2/2}$, at the nodes of an $80$-point Gauss-Hermite quadrature, so that every scalar product becomes a weighted sum whose integrand is a polynomial times $e^{c\sigma x}$ with $|c|\le2\pi$. The quadrature is exact up to terms that are negligible at double precision, and the results change by less than $10^{-15}$ when $120$ or $200$ nodes are used. The modular maps act pointwise on the nodes, $\delta^{\pm1/4}$ as multiplication by $e^{\pm\pi\sigma x/2}$ and $j$ as complex conjugation combined with the reflection of the symmetric node set. The $8N$ real coefficients of the four vectors of each party in the basis of $\Rr_N$ are optimized with the BFGS method, the exact gradient of $\BCHSH$ being obtained by automatic differentiation, from $16$ to $30$ random starting points. At the optimum the residuals of the localization conditions, $sf=f$ for Alice and $s^\dagger g=-g$ for Bob, are of order $10^{-15}$, the real parts of the overlaps between the two parties are of order $10^{-17}$, and the imaginary part of $\BCHSH$ is below $10^{-18}$. The explicit configuration of Sec.~\ref{sec:gaussian} uses the compressions of $\omega^2$, in units of $\sigma^2$, to the even and odd parts of $\Rr_6$, whose eigenvalues are $0.19016$, $1.78449$, $5.52534$ and $0.66633$, $2.80078$, $7.03290$. The compact seeds of Secs.~\ref{sec:compact} and~\ref{sec:pauli} require only the one-dimensional integrals $I(a)$ and $\mu_2$, which we compute by adaptive quadrature. We have verified the closed forms \eqref{eq:corr-compact}, \eqref{eq:T} and \eqref{eq:Bopt} against a direct evaluation of \eqref{eq:wick} on a grid of $4\times10^5$ points in $\omega$, with agreement to $10^{-15}$ for the real parts of all internal and mixed overlaps and to six digits for the Bell parameters.

\bibliography{refs}

\end{document}